\documentclass[a4]{article}
\usepackage[utf8]{inputenc}
\usepackage[english]{babel}
\usepackage{amssymb,mathtools, bm}
\usepackage[left=2.5cm, right=2.5cm, top=2.5cm, bottom=2.5cm]{geometry}
\usepackage{graphics, float}
\usepackage{hyperref, authblk, caption}
\usepackage{appendix}
\usepackage{todonotes}

\newcommand{\trans}[4]{
	\left(
		\begin{matrix} #1 & #2 \\ #3 & #4 \end{matrix}
	\right)
	}

\newcommand{\colvec}[2]{
	\left(
		\begin{matrix} #1 \\ #2 \end{matrix}
	\right)
	}

\newcommand{\I}{\mathrm{i}}
\newcommand{\E}{\mathrm{e}}

\title{Double-slit interferometry as a lossy beam splitter}
\date{}
\author[1,*]{Simanraj Sadana}
\author[1,2,$\ddagger$]{Barry C. Sanders}
\author[1,$\dagger$]{Urbasi Sinha}

\affil[1]{Light and Matter Physics Group, Raman Research Institute, Bangalore 560080, India}
\affil[2]{Institute for Quantum Science and Technology,
	University of Calgary, Alberta T2N 1N4, Canada}

\begin{document}

\maketitle
\noindent*Email: simanraj@rri.res.in \\
$\ddagger$Email: sandersb@ucalgary.ca \\
$\dagger$Email: usinha@rri.res.in
\begin{abstract}
We cast diffraction-based interferometry in the framework of post-selected unitary description towards enabling it as a platform for quantum information processing. We express slit-diffraction as an infinite-dimensional transformation and truncate it to a finite-dimensional transfer matrix by post-selecting modes. Using such a framework with classical fields we show that
a customized double-slit setup is effectively a lossy beam splitter in a post-selected sense. Diffraction optics provides a robust alternative to conventional multi-beam interferometry with scope for miniaturization, and also has applications in matter wave interferometry. In this work, the classical treatment of slit-diffraction sets the stage for quantization of fields and implementing higher-dimensional quantum information processing like that done with other platforms such as orbital angular momentum.
\end{abstract}

\section{Introduction}\label{sec:introduction}
Linear optical quantum information processing (QIP) \cite{Knill:2001aa,kok2007} has a mathematical representation in the form of finite-dimensional unitary transfer matrices operating on a Hilbert space of vectors that represent qubits/qudits \cite{Genovese:2008:1936-6612:153, padua2006}. The qubits are usually encoded in the polarization degree of freedom of a single photon, and optical components like beam splitters \cite{saleh1991fundamentals, gerry2005introductory, Garc_a_Escart_n_2011} and phase-shifters are used to implement the unitary transformations on them. For higher-dimensional QIP, systems such as orbital angular momentum \cite{Garc_a_Escart_n_2011, Mair:2001aa, Molina-Terriza:2007aa, calvo2007, Malik:12} of photons are used. We map diffraction optics over a finite-dimensional unitary representation and connect it to qubit/qudit processing.

The novel interpretation of slit-diffraction that we present here sets the stage for extending the scope of application of diffraction interferometry to modern problems like higher-dimensional information processing. Such a formalism 
provides an alternative to the implementation of higher-dimensional QIP using the orbital angular momentum (OAM) of light \cite{Garc_a_Escart_n_2011}. Slit-diffraction based optical interferometers can be used to construct qudits encoded in spatial modes \cite{xiaoDow2017,Ghosh:18}, with robustness, unlike in the case of OAM based qudits which have practical limitations in state-preparation and state-readability \cite{Garc_a_Escart_n_2011}. Another potential advantage of the multi-slit-diffraction-based interferometer is scalability of table-top experiments. Moreover, a finite-dimensional unitary description of diffraction also has applications in the field of matter-wave interferometry \cite{THOMSON:1927aa, davisson1927, RAUCH1974369, rauch2015neutron, keith1988, keith1991}.

We deal with diffraction of classical fields and show a formalism in which slit-diffraction is represented as a finite-dimensional unitary transfer matrix \cite{born2013principles} (in the post-selected sense). We project the  three-dimensional solutions of the Helmholtz equation \cite{born2013principles, jackson1977classical} on two-dimensional imaginary planes and call these projections slices. The propagation and diffraction of the fields is expressed as a slice-to-slice map as one goes from one slice to another from the sources to the detectors through the slits. By choosing an appropriate basis for the slices, we get an infinite-dimensional transfer matrix representation of such a map. The transfer matrix is reduced to an effective finite-dimensional matrix by post-selecting a finite number of basis elements on the slices as post-selected modes. We show that such a truncated matrix is in general not unitary because of the losses in diffraction, and that the underlying unitary transfer matrix can be revealed by performing a polar decomposition \cite{hall2015lie} on the effective transfer matrix separating it into unitary and lossy (Hermitian) components.

Using the post-selected unitary transfer matrix formalism of diffraction, we show that a customized double-slit setup is effectively a lossy beam splitter in the classical regime. A cubic beam splitter is a two-input-two-output optical device that has a $2\times 2$ unitary transfer matrix that transforms the fields entering its input ports to the fields exiting its output ports \cite{saleh1991fundamentals, gerry2005introductory}. This 4-port device, along with a phase-shifter which is a 2-port device that imparts a phase, serves as the building blocks of any $N$-channel interferometer \cite{vilenkin1968special, miller1968lie,talman1968special,reck1994experimental,bouland}. The novelty and importance of our work lies in connecting one of the most elegant and fundamental experiments in scientific history, i.e., double-slit-diffraction with other types of interferometries which are used to solve some of the most important problems in modern physics, like QIP.

To verify the beam splitter like behaviour of the double-slit setup, we compare the correlation of the classical outputs with
that of the cubic beam splitter. Moreover, by concatenation of two such double-slit based beam splitters and using a phase-shifter, we construct an effective Mach-Zehnder interferometer \cite{hariharan2010basics}. The two-dimensional transfer matrix representation of double-slit-diffraction validates the formalism and allows us to extend to higher-dimensional system and find a transfer matrix representation for the same. Here we show such an application by finding the transfer matrix for a triple-slit system, demonstrating the way to extend the formalism from two slits to a higher number of slits.

\section{Background}\label{sec:background}
The transfer matrix representation of diffraction presented in this paper uses concepts from classical optics (Helmholtz equation), signal processing (wavelets) as well as linear algebra. A brief discussion of these concepts and their relevance in this work is presented in this section. 

\subsection{The Helmholtz equation and Hilbert space}\label{subsec:helmholtz}
To represent diffraction as a transformation in a Hilbert space, we use solutions of the Helmholtz equation \cite{born2013principles,jackson1977classical,mandel1995optical}. The Helmholtz equation is a self-adjoint linear partial differential
equation. Therefore, its solutions or fields have a vector in a Hilbert
space associated with them. Moreover, the projections of the
three-dimensional (3D) fields on two-dimensional (2D) planes, say the \(xy\)
plane also form a Hilbert space. It should be noted that there is a difference between
two-dimensional fields and the projections of 3D fields on 2D
planes. The 2D projections are referred to as \emph{slices} of the 3D fields.

Diffraction of light is understood by solving the Maxwell's equations \cite{born2013principles,jackson1977classical,mandel1995optical}, specifically the wave equation, with appropriate boundary conditions. Generally, the time-dependence of solutions (or fields) is considered harmonic, i.e., of the form $\mathrm{e}^{\mathrm{i} \omega t}$, where $\omega$ is the angular frequency. Consequently, the wave equation reduces to the time-independent Helmholtz equation \cite{born2013principles,jackson1977classical,mandel1995optical}. For a source $\rho(\bm{r})$ in a volume $\mathcal{V}$ enclosed by a surface $\partial \mathcal{V}$ on which the boundary condition is specified, the most general solution of the Helmholtz equation is,
\begin{align}
	\label{eq:genSolHelmholtz}
	E(\bm{r}) = \iiint\limits_{\mathcal{V}} \mathrm{d}^3 \bm{r'}~ G(\bm{r},\bm{r}') \rho(\bm{r}') + \iint\limits_{\partial\mathcal{V}} \mathrm{d}^2 \bm{r}'~ \hat{n}(\bm{r}') \cdot \left( E(\bm{r}') \nabla' G(\bm{r},\bm{r}') - G(\bm{r},\bm{r}') \nabla' E(\bm{r}') \right),
\end{align}
where $\hat{n}(\bm{r}')$ is unit normal to the surface and $G(\bm{r},\bm{r}')$ is the Green's function for the Helmholtz equation. 

We apply the Fraunhofer approximation to Eq.~(\ref{eq:genSolHelmholtz}) (see appendix \ref{append:Fraunhofer} for details) to find solutions of the Helmholtz equation and project them onto the $xy$ plane by fixing $z$. Using the surface term in Eq.~(\ref{eq:genSolHelmholtz}) we find a slice-to-slice map (see appendix \ref{append:slice2slice}). To get a matrix representation of the slice-to-slice map, we represent each slice as a column vector in a suitable basis on that slice.

One set of orthonormal vectors that span the Hilbert space on a slice, can be found by finding the eigensolutions of the homogeneous Helmholtz equation, with the appropriate boundary conditions \cite{jackson1977classical}, and using them as modes. For example, the eigensolutions are standing sinusoidal waves if the boundary condition is, say reflective. However, these modes are not localized and thus unsuitable for finite number of detectors with a given size. Therefore, we choose a basis of two-dimensional functions with compact support, that spans the Hilbert space on a slice.

\subsection{Haar wavelets}\label{subsec:wavelets}
We use Haar wavelets and scaling function \cite{mallat1999wavelet,kaiser2010friendly} to construct orthonormal basis for a slice (projections of fields on a plane, see \S\ref{subsec:helmholtz}). 
Compact support of the wavelets makes them suitable modes for detectors that have finite size. The orthogonality of the wavelets ensures that there is no overlap between measurements by two detectors. For a detector with square-shaped window, the two-dimensional Haar wavelets \cite{Haar1910} are chosen.

\begin{figure}[H]
	\centering
	\begin{minipage}{0.45\textwidth}
		\centering
		\includegraphics[width=1\textwidth]{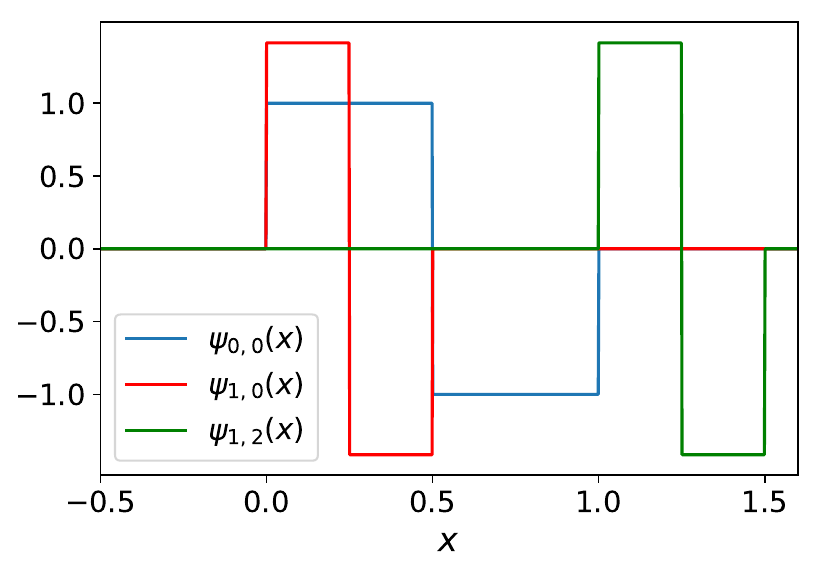}
		\caption{Examples of dilated and translated Haar wavelets defined in Eq.~(\ref{eq:haarWavelets}). The plots of the functions clearly show the orthogonality of the functions with respect to the overlap integral in Eq.~(\ref{eq:waveletOrthogonality}) as the inner product.}
		\label{fig:dilTransMother}
	\end{minipage}\hfill
	\begin{minipage}{0.45\textwidth}
		\centering
		\includegraphics[width=1\textwidth]{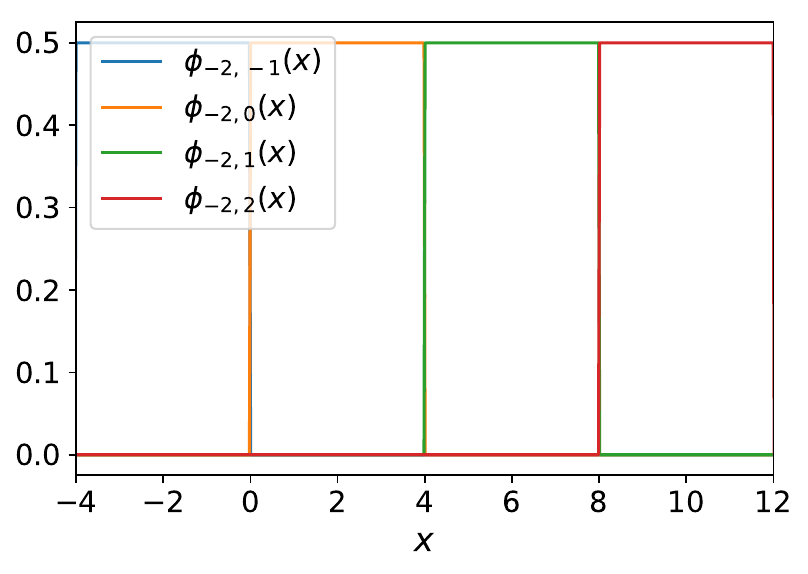}
		\caption{Examples of translated Haar scaling function at a particular scale $j=-2$. At this scale the translated Haar scaling functions do not have any overlap with each other, making them orthogonal to each other.}
		\label{fig:dilTransFather}
	\end{minipage}

	\label{fig:dilTransWavelets}
\end{figure}

Wavelets are square-integrable functions with compact support over a finite interval. The simplest example is the Haar wavelet \cite{mallat1999wavelet,kaiser2010friendly,Haar1910}, which is defined by its wavelet function $\psi$ (or mother function) and a scaling function $\phi$ (or father function)
\begin{align}
	\label{eq:haarMother}
	\psi(x) \coloneqq &\sqcap \left(2 \left(x-\frac{1}{4}\right)\right)-\sqcap \left(2\left(x-\frac{3}{4}\right)\right),\\ 
	\label{eq:haarFather}
	\phi(x) \coloneqq &\sqcap \left(x-1/2 \right),
\end{align}
respectively, where $\sqcap$ is the box function. These functions are dilated and translated to create other Haar wavelets and scaling functions,
\begin{align}
	\label{eq:haarWavelets}
	\psi_{m,n}(x) \coloneqq &\frac{1}{\sqrt{2^{-m}}} \psi \left(\frac{x}{2^{-m}}- n\right),\\
	\label{eq:haarScalings}
	\phi_{j,k}(x) \coloneqq  &\frac{1}{\sqrt{2^{-j}}} \phi \left(\frac{x}{2^{-j}}- k\right),
\end{align}
respectively, where $m$ and $j$ are dilation parameters whereas $n$ and $k$ are translation parameters, and all take integer values. Examples of dilated and translated wavelet and scaling functions are shown in Figs.~\ref{fig:dilTransMother}
and~\ref{fig:dilTransFather} respectively.

The Haar wavelets are orthogonal with respect to the inner product
\begin{align}
	\label{eq:waveletOrthogonality}
	\left\langle \psi_{m,n}, \psi_{m',n'} \right\rangle \coloneqq \int\limits_{-\infty}^{\infty} \mathrm{d} x~ \psi_{m,n}^*(x) \psi_{m',n'}(x) = \delta_{m,m'} \delta_{n,n'},
\end{align}
and the Haar scaling functions are also orthogonal, i.e., 
\begin{align}
	\label{eq:scalingOrthogonality}
	\left\langle \phi_{j_0,k}, \phi_{j_0, k'} \right\rangle = \delta_{k,k'},
\end{align}
at a particular scale, say, $j = j_0$ (see Fig.~\ref{fig:dilTransFather}). In signal processing, any signal $f(x)$ can be decomposed into Haar wavelets and scaling function of a particular scale $j_0$  \cite{mallat1999wavelet,kaiser2010friendly,Haar1910} as
\begin{align}
	\label{eq:haarDecomposition}
	f(x) = \sum\limits_{k=-\infty}^{\infty} a_k \phi_{j_0,k}(x) + \sum\limits_{m \geq j_0} \sum\limits_{n = -\infty }^{\infty} b_{m,n} \psi_{m,n}(x),
\end{align}
where the scaling function is equivalent to a low-pass filter and the wavelet functions are equivalent to band-pass filters. 

By constructing a basis using Haar wavelets on each slice, we represent the projected field as a column vector in that basis, and express the slice-to-slice map as a transfer matrix between two slices. To validate our formalism we use this approach to show that a double-slit system is effectively a lossy beam splitter and verify it by studying the correlation of the outputs of the double-slit setup and compare it with that of a cubic beam splitter. 
\subsection{Beam splitter and its transfer matrix}\label{subsec:beamsplitter}
A beam-splitter is a ubiquitous two-input-two-output component in interferometry. In optics, a beam splitter is commonly in the form of a glass cube, half-silvered mirror or fibre based, which have two input and two output modes corresponding to each of their ports. In a 50:50 cubic beam splitter, for example, the modes are the $k$-vectors corresponding to the plane wave entering each of its ports, forming a basis to represent the inputs and outputs as two-dimensional column vectors in a Hilbert space. In such a representation, the beam splitter transformation has a two-dimensional transfer matrix representation \cite{saleh1991fundamentals,gerry2005introductory,loudon2000quantum}, denoted here by
\begin{align}
	\label{eq:bsMatrix}
	U_{\mathrm{BS}} = \frac{1}{\sqrt{2}} \trans{1}{\mathrm{i}}{\mathrm{i}}{1},
\end{align}
where each row corresponds to the superposition of the two input modes to form the outputs, and complex elements of the matrix denote the phase-shift introduced in each input. 

In general, if the source of light does not emit in a single mode (say, a divergent beam), the vector representation of the inputs and outputs can be infinite-dimensional, yielding an infinite-dimensional transfer matrix of the beam splitter. In such a case, two suitable input and two output modes can be post-selected to reduce the infinite-dimensional transfer matrix to a post-selected $2 \times 2$ transfer matrix as in Eq.~(\ref{eq:bsMatrix}). 

A consequence of such a transformation is that the outputs of the beam splitter are correlated, as discussed in the next subsection. The correlation of the outputs, as a function of a parameter that distinguishes the inputs, is a signature of a beam splitter. We use this signature to verify the claim that a double-slit setup is effectively a beam splitter.

\subsection{Correlated outputs of a beam splitter}
In semi-classical theory of photo-detection \cite{Mandel_1958,Mandel_1964,mandel1995optical}, the probability of coincident photo-detections is proportional to the intensity-intensity cross-correlation of light in the post-selected modes, falling on the detectors. Such a correlation of the outputs of a 50:50 beam splitter plotted as a function of some distinguishability parameter shows a dip \cite{Ou90,Ou07,GRA86} for identical pulses (or photon states in the quantum regime \cite{HOM87}) at the input ports. A parameter, say, time-delay between the input pulses, distinguishes the otherwise identical input pulses. The correlation depends on the shape of the input pulses and the fluctuations in the light field. Specifically, if the fluctuation is uniform, the correlation shows a dip of 50\% as the distinguishability parameter (like time-delay) approaches zero. A brief discussion of this concept is in appendix \ref{append:corr} and a detailed analysis of this phenomenon is presented in \cite{sadana2018near}.

Combination of the above concepts have been used to cast diffraction optics in the framework of post-selected unitary description. Consequently, the mathematical framework of diffraction optics becomes at par with that of other types of interferometry. The classical treatment outlined in the coming sections sets the stage for quantization of fields enabling diffraction optics as an alternative platform for QIP.

\section{Approach and method}
\label{sec:approach}
Here we discuss the approach towards the transfer matrix formalism of diffraction using a double-slit setup as an example, and then extend its application to find the transfer matrix of a triple-slit setup. We discuss the slice modes using Haar functions and the column vector representation of the slices. Then we truncate the dimensionality by choosing certain Haar functions as post-selected in input/output modes, and finding an effective $2 \times 2$ transfer matrix for double-slit-diffraction, showing that it behaves like a lossy beam splitter.
We verify the efficacy of the double-slit beam splitter by studying the correlation of the outputs and, by making an MZI by concatenating two double-slit beam splitters. Finally, we use the transfer matrix formalism to find the transfer matrix of a triple-slit system demonstrating its application to higher-dimensional systems.
\subsection{The slice modes}
\label{subsection:sliceModes}
As discussed in \S\ref{subsec:helmholtz}, 
non-local eigenfunctions
of the Helmholtz equation do not make suitable modes for detectors with
finite-sized windows. Haar functions (\S\ref{subsec:wavelets}) on the other hand, have
compact support over a given interval and therefore two-dimensional Haar
functions make suitable modes for the square-shaped detector windows.
The Haar wavelets and the Haar scaling functions, however, form an
overcomplete set of orthonormal functions \cite{kaiser2010friendly,mallat1999wavelet}.

\begin{figure}[H]
	\centering
	\includegraphics[width=0.3\textwidth]{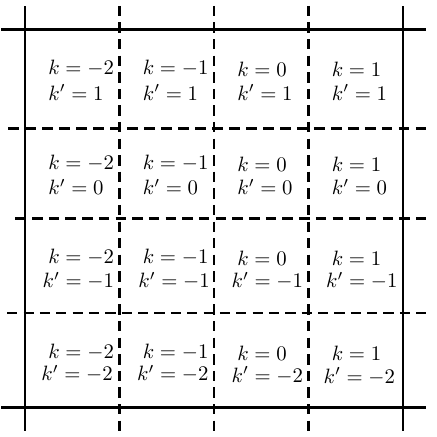}
	\caption{An example of how a plane at some $z$ can be segmented into non-overlapping square patches indexed by two integers $k$ and $k'$. The width of each patch is equal to the width of the available detector. To know the entire slice at $z$, all the patches must be considered. But usually, there are a finite number of detectors so that only a few patches can be covered. In that case, only those Haar wavelets and scaling functions are considered which have a compact support on the considered patches. }
	\label{fig:squarePatches}
\end{figure}

To remove the overcompleteness, we divide each slice into
non-overlapping square patches,
each with side-length equal $w$, as shown in
Fig.~\ref{fig:squarePatches}. The square patches are labelled using
two indices \(k\) and \(k'\) which take integer values. 
\noindent Each square
patch supports a countably infinite set of Haar wavelets that fall
entirely within the patch. Together with the Haar scaling function that
covers the square patch, all the supported Haar wavelets form a basis
for any function that has support over the patch. The first element of
this basis is the Haar scaling function that covers the entire patch, i.e.,
\begin{align}
    \label{eq:supportedScaling}
    g_1(x,y;z_1,j_0,k,k') \coloneqq \phi_{j_0,k}(x)~ \phi_{j_0,k'}(y - j_0),
\end{align}
where the dilation parameter of the scaling function, i.e., \(j_0\) is
set so that $g_1(x,y;z_1,j_0,k,k')$ covers the entire patch (see Eq.~(\ref{eq:haarScalings})), and $z= z_1$ is the plane on which the slice is considered.
The other elements of the basis are all Haar wavelets with
compact support over the square patch, i.e.,
\begin{align}
    g_\imath(x,y;z_1,j_0,k,k') := \psi _{m,n}(x) \psi_{m'n'}(y - j_0)~  \forall~ \imath>1~\in~\mathbb{Z}^+, \label{eq:supportedWavelet}
\end{align}
where \(\mathbb{Z}^+\) is the set of positive integers. The subscript $\imath$
is a meta-index for \(m\), \(n\), \(m'\) and \(n'\), and
\begin{align}
	&m \geq j_0, \\
	2^{m-j_0} k  \leq &n < 2^{m-j_0}(k+1), \\
	&m' \geq j_0, \\
	2^{m'-j'_0} k'  \leq &n' < 2^{m'-j'_0}(k'+1),
\end{align}
where the ranges ensure that all the wavelets have compact support over
the square patch chosen. If such Haar functions for all the square
patches are combined, the slice can be resolved in terms of these
functions using Eq.~(\ref{eq:haarDecomposition}).

\subsection{The double-slit setup}
\label{sec:doubleSlitSetup}
\begin{figure}[H]
\centering
\includegraphics[width=0.5\textwidth]{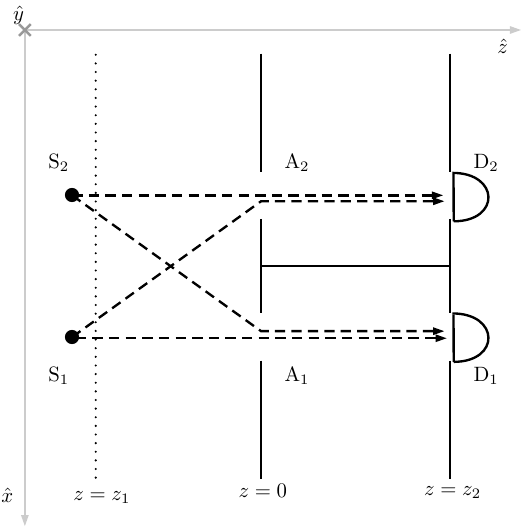}
\caption{Schematic of the double-slit setup considered in this paper. Two point sources $\mathrm{S}_1$ and $\mathrm{S}_2$ emanate monochromatic linearly polarized light with harmonic time-dependence. The imaginary plane at $z=z_1$ (represented by a dotted line) is for the input slice. Two slits $\mathrm{A}_1$ and $\mathrm{A}_2$ are placed at $z=0$ where each slit is aligned center-to-center with one of the sources. A second pair of slits $\mathrm{D}_1$ and $\mathrm{D}_2$ are placed at $z=z_2$ where each port is aligned center-to-center with one of the source. Behind each of these slits is a square-faced detector which measures the integrated intensity of the light falling on it. The plane $z=z_2$ is also for projecting the output slice. A perfectly absorbing barrier runs between $z=0$ and $z=z_2$ that prevents the field from slit $\mathrm{A}_1$ (A$_2$) from reaching port $\mathrm{D}_2$ (D$_1$). The dashed arrows represent the ray approximations of the fields from the sources to the detectors.}
	\label{fig:doubleSlitSetup}
\end{figure}

We elaborate on the slice modes concept using a customized double-slit setup with two sources and two detectors as shown in Fig.~\ref{fig:doubleSlitSetup}, where $\hat{\bm{y}}$  extends into the plane of the paper. The slits are parallel to the $xy$ plane and so are the sources and the detectors at different values of $z$. The width of the apertures and other distances are chosen such that far-field approximations can be applied to solutions of the Helmholtz equation. A perfectly absorbing barrier is added that runs along the $z$ direction and separates a slit from the detector across the barrier. The purpose of the barrier is to prevent the high diffraction orders \cite{born2013principles} from reaching the detectors and also to isolate one detector from another to avoid an overlap of fields between the two.

The sources $\mathrm{S_1}$ and $\mathrm{S_2}$ are monochromatic point-like sources (practically a spherical source with diameter $\sim\lambda$) emanating linearly polarized light as spherical waves ($Y_0^0(\theta, \phi)$ spherical harmonic \cite{jackson1977classical}) with wavelength $\lambda$. They are placed at $\bm{r}_\mathrm{S_1} = \left(d/2,0,-L\right)$ and $\bm{r}_\mathrm{S_2} = \left(-d/2,0,-L\right)$ respectively, where $d=20 \lambda$ is the distance between the two sources and $L=800\lambda$ is the distance between the sources and the slit plane along the $z$ direction. Without loss of generality, we choose $\lambda = 1$.

In the far-field regime \cite{born2013principles}, these sources can be approximated by Dirac-delta functions $\delta^3(\bm{r} - \bm{r}_{\mathrm{S}_1})$ and $\delta^3(\bm{r} - \bm{r}_{\mathrm{S}_2})$. We multiply the source-term with a factor of $10^5$ so that the simulation results do not suffer precision errors. As the sources are linearly polarized, the field from source S$_i$ at points far from the slits, before diffraction, can be found by solving the Helmholtz equation for scalar fields and can be approximated by
\begin{align}
\label{eq:sourceField}
	E^{(i)}(\bm{r}) \approx& G(\bm{r}, \bm{r}_{\mathrm{S}_i}),
\end{align}
where the use of the scalar equation is justified because the polarizations of field from both the sources are collinear. Note that the approximation in Eq.~(\ref{eq:sourceField}) is valid only because the slit plane is far enough from the plane at $z_1$, so that the surface-effects are negligible.

Two square-shaped slits $\mathrm{A_1}$ and $\mathrm{A_2}$,
each with side-length $w=4\lambda$ are placed at $z=0$ with the positions of their centers $\bm{r}_\mathrm{A_1} = \left(d/2,0,0\right)$ and $\bm{r}_\mathrm{A_2} = \left(-d/2,0,0\right)$ respectively, and therefore aligned with the respective sources. In the Fraunhofer regime, the diffracted field $E^{(i)}_j(\bm{r})$ from source S$_i$ through slit A$_j$, is calculated by simplifying the surface term in Eq.~(\ref{eq:genSolHelmholtz}) by applying the appropriate approximations (see appendix \ref{append:Fraunhofer}).
Note that,
due to the opaque barrier between the two detectors, detector port D$_k$ is blocked from the field $E^{(i)}_j(\bm{r})$ if $k \neq j$.

Another pair of square-shaped slits $\mathrm{D_1}$ and $\mathrm{D_2}$, aligned with $\mathrm{A_1}$ and $\mathrm{A_2}$ respectively, are placed at $z=L$, behind each of which is a square-law detector (that measures the integrated squared magnitude of fields) whose window is of the same shape and size as those of the slits. 
The detector is 100\% efficient for light of wavelength $\lambda$. The detectors could have been placed without the second pair of slits which play a role only when two such double-slit setups are concatenated to construct interferometers. 

Two imaginary planes are at $z_1 = -0.9 L$ and $z_2 = L$ on which the input and output slices are considered respectively, for the double-slit. The input slice is not placed at $-L$ where the sources are, as the solution of the Helmholtz equation diverges at the sources.
The 2D wavelets are used as basis functions (as discussed in \S\ref{subsection:sliceModes}) on each of these slices so that each slice can be represented as a column vector. By choosing the input and the output slices on either side of the slits, a transfer matrix mapping the input slice to the output slice can be calculated. However, the slice-to-slice map can also be done in a continuous manner where a propagator sequentially maps one slice to another very close to it, gradually moving forward in the $z$ direction. Such a map is constructed using the surface term of the formal solution of the Helmholtz equation (see Eq.~(\ref{eq:genSolHelmholtz})), which is discussed in appendix \ref{append:slice2slice}. However, for the purpose of showing that a double-slit is effectively a beam splitter, a direct transfer matrix between the input and output slices suffices.

\subsection{The post-selected slice modes}\label{subsec:post-selectedModes}
The two detector-windows in the double-slit setup cover only two of the square
patches. Consequently, they do not intercept the entire slice, but only
a portion of it. Nevertheless, each detector window supports a countably
infinite number of Haar functions. We justify the post-selection of two input and two output modes.

\subsubsection{Output modes}\label{header-n38}
As the width of the detector is $w=4$, we set $j_0=-2$ so that the Haar scaling function covers the entire patch. According to the positions of the detectors in Fig.~\ref{fig:doubleSlitSetup}, the square patch occupied by the detector at port D\(_1\) is the one with
indices \(k=2,~k'=0\). Similarly, the patch covered by the detector at port
D\(_2\) is the one with \(k=-3,~ k'=0\). From the infinite set of Haar
functions supported by the square patches, two have to be post-selected.
There are two ways to achieve that.

One way is to design a detector that responds to the projection of light
on a particular Haar wavelet or scaling function. Although possible in
principle, making such a detector is practically challenging because of
the jump discontinuities in the Haar wavelet functions. Another and more
tractable approach is to construct the double-slit setup in such a way
that most of the light intercepted by the detectors has projection on a
single Haar wavelet or the scaling function, which becomes a detector
mode. Consequently, even if the detector is multimode, the detection is
in single mode. The latter is the case with the double-slit setup
considered in this work.

\begin{figure}[H]
	\centering
	\begin{minipage}{0.45\textwidth}
		\includegraphics[width=1\textwidth]{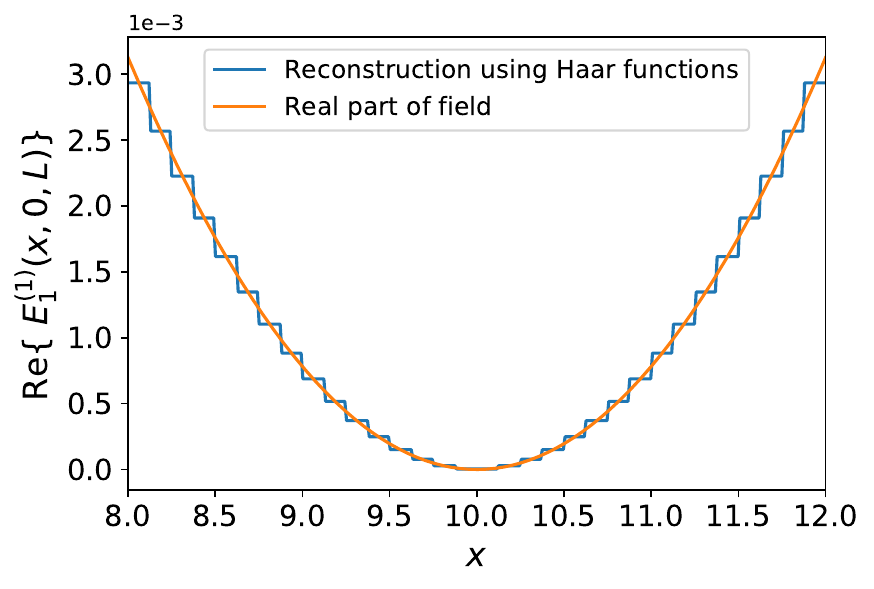}
		\caption{Reconstruction of the real part of $E^{(1)}_1(x,0,z_2)$ using the wavelets supported over the patch $k=2,~k'=0$, using Eq.~(\ref{eq:EExpanded}). The values of dilation parameter $m$ for the Haar wavelets (as in Eq.~(\ref{eq:haarWavelets})) is taken from $-2$ to $2$ so that the Haar wavelets are visible. A finer reconstruction can be done by taking $m$ upto higher values. For $m$ upto 6, the reconstruction is almost perfect.}
		\label{fig:reconstE1Re}
	\end{minipage}\hfill
	\begin{minipage}{0.48\textwidth}
		\includegraphics[width=1\textwidth]{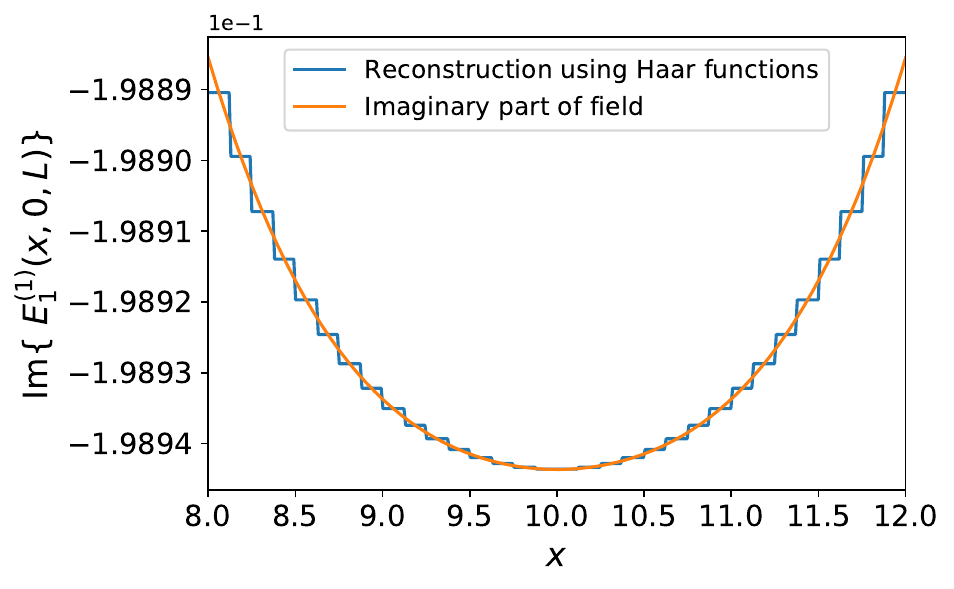}
		\caption{Reconstruction of the imaginary part of $E^{(1)}_1(x,0,z_2)$ using the wavelets supported over the patch $k=2,~k'=0$, using Eq.~(\ref{eq:EExpanded}). The values of dilation parameter $m$ for the Haar wavelets (as in Eq.~(\ref{eq:haarWavelets})) is taken from $-2$ to $2$ so that the Haar wavelets are visible. A finer reconstruction can be done by taking $m$ upto higher values. For $m$ upto 6, the reconstruction is almost perfect.}
		\label{fig:reconstE1Im}
	\end{minipage}
\end{figure}

To find such modes, the diffracted fields intercepted by the detector
windows (ports D\(_1\) and D\(_2\)) are resolved in terms of the
corresponding Haar functions. For example, the field
\(E^{(1)}_{1}(x,y;z_2)\) can be expanded as
\begin{align}
	\label{eq:EExpanded}
	E^{(1)}_{1}(x,y;z_2) = \sum\limits_{\imath=1}^{\infty} A_\imath(z_2)~ g_\imath(x,y;z_2,-2,2,0),
\end{align}
where \(A_\imath(z_2)\) are the projections of the field on the corresponding
Haar function. For convenient visualization, the field at \(y=0\), i.e.,
\(E^{(1)}_1(x,0;z_2)\) is shown in Figs.~\ref{fig:reconstE1Re} and \ref{fig:reconstE1Im}.
Although it takes more than one basis function to capture the fine
features of the field, most of the power of the light resides in just
one mode, i.e., \(g_1(x,y;z_2,-2,2,0)\). The proof of this fact is in Table \ref{tab:ratioOfLight}, where the total power intercepted by port D\(_1\) is compared
with the total power in the projection on \(g_1(x,y;z_2,-2,2,0)\). The
calculation shows that about \(99.9956\%\) of the total power is in the
said projection. The field \(E^{(2)}_{1}(x,y;z_2)\) also has most of the
power in this mode. Therefore, we ignore all the other
Haar functions which have negligible contribution to the field.
Similarly, for detector port D\(_2\), the dominant contribution is from
\(g_1(x,y;z_2,-2,-3,0)\). Henceforth, two post-selected output modes are
\begin{align}
	\label{eq:PSOutMode1}
	e_1(x,y;z_2) \coloneqq & g_1(x,y;z_2,-2,2,0),\\
	\label{eq:PSOutMode2}
	e_2(x,y;z_2) \coloneqq & g_1(x,y;z_2,-2,-3,0),
\end{align}
which reduce the infinite-dimensional representation of the slice at
\(z=z_2\) to a two-dimensional column vector
	\begin{align}
	\label{eq:outputFromSi}
	\bm{Y}^{(i)}(z_2) = \colvec{\left\langle e_1(z_2),  E^{(i)}_1(z_2) \right\rangle}{\left\langle e_2(z_2),  E^{(i)}_2(z_2) \right\rangle},
	\end{align}
where \( E^{(i)}_j(x,y;z_2)\) is the field from source S\(_i\) diffracted by
slit A\(_j\) and therefore intercepted by the detector port D\(_j\), and
\begin{align}
    \left\langle e_1(z_2),  E^{(i)}_1(z_2) \right\rangle =& \int\limits_{-\infty}^\infty \mathrm{d}x \int\limits_{-\infty}^\infty \mathrm{d}y~ e_1^*(x,y;z_2)~ E^{(i)}_1(x,y;z_2)
\end{align}
and similarly for the second entry in the column vector.
{
	\renewcommand{\arraystretch}{3}
	\begin{table}[H]
		\centering
		\begin{tabular}{|c|c|c|}
			\hline
			\textbf{Integrated intensity intercepted} & \textbf{Integrated projections on chosen modes} & \textbf{Ratio}\\
			\hline
			$\iint\limits_{\mathrm{D}_1} \mathrm{d} x~ \mathrm{d} y~ \left|\bar{E}^{(1)}_1(x,y;z_2) \right|^2 = 0.633115$ & $\iint\limits_{\mathrm{D}_1} \mathrm{d} x~ \mathrm{d} y~ \left| \left\langle e_1(z_2), \bar{E}^{(1)}_1(z_2) \right\rangle e_1(x,y;z_2)\right|^2  = 0.633087$ & $99.9956\%$\\
			\hline
			$\iint\limits_{\mathrm{D}_2} \mathrm{d} x~ \mathrm{d} y~ \left|\bar{E}^{(1)}_2(x,y;z_2) \right|^2 = 0.61200$ & $\iint\limits_{\mathrm{D}_2} \mathrm{d} x~ \mathrm{d} y~ \left| \left\langle e_2(z_2), \bar{E}^{(1)}_2(z_2) \right\rangle e_2(x,y;z_2)\right|^2 = 0.611971$ & $99.9952\%$\\
			\hline
			$\iint\limits_{\mathrm{D}_1} \mathrm{d} x~ \mathrm{d} y~ \left|\bar{E}^{(2)}_1(x,y;z_2) \right|^2 = 0.61200$ & $\iint\limits_{\mathrm{D}_1} \mathrm{d} x~ \mathrm{d} y~ \left| \left\langle e_1(z_2), \bar{E}^{(2)}_1(z_2) \right\rangle e_1(x,y;z_2)\right|^2 = 0.611971$ & $99.9952\%$\\
			\hline
			$\iint\limits_{\mathrm{D}_2} \mathrm{d} x~ \mathrm{d} y~ \left|\bar{E}^{(2)}_2(x,y;z_2) \right|^2 = 0.633115$ & $\iint\limits_{\mathrm{D}_2} \mathrm{d} x~ \mathrm{d} y~ \left| \left\langle e_2(z_2), \bar{E}^{(2)}_2(z_2) \right\rangle e_2(x,y;z_2)\right|^2 = 0.633087$ & $99.9956\%$\\
			\hline
		\end{tabular}
		\caption{A comparison of the total integrated intensity detected by the detectors and the square of the magnitudes of the projections of the fields on the chosen modes. The ratios show that most of the light intercepted by the detectors are in the chosen modes as in Eqs.~(\ref{eq:PSOutMode1}) and (\ref{eq:PSOutMode2}). Therefore the choice of modes is justified.}
		\label{tab:ratioOfLight}
	\end{table}
}

\subsubsection{Input modes}\label{header-n48}
Each source in the double-slit setup (Fig.~\ref{fig:doubleSlitSetup}) emits light in a particular mode, for example
spherical mode (approximately). The projection of the source modes on the slice at \(z=z_1\)
are projected on the Haar functions on that plane. The input ports
chosen are square patches on the plane at \(z_1\) centered at the same
positions on the plane as the output ports are placed on plane at
\(z_2\). Moreover, the post-selected input modes are similar to those
chosen for the output, i.e.,
\begin{align}
	\label{eq:PSInputMode1}
    e_1(x,y;z_1) \coloneqq & g_1(x,y;z_1,-2,2,0),\\
    \label{eq:PSInputMode2}
    e_2(x,y;z_1) \coloneqq & g_1(x,y;z_1,-2,-3,0),
\end{align}
with \(z_1\) denoting that the modes are for a slice on plane \(z=z_1\).
Note that the input and output modes are distinguished using the
parameter that denotes the plane on which the slice is, i.e., \(z\).
Therefore, the post-selected two-dimensional vector representation of
the input, i.e., slice at \(z=z_1\)
	\begin{align}
	\label{eq:inputFromSi}
	\bm{X}^{(i)}(z_1) = \colvec{\left\langle e_1(z_1),  E^{(i)}_1(z_1) \right\rangle}{\left\langle e_2(z_1),  E^{(i)}_2(z_1) \right\rangle},
	\end{align}
where the superscript denotes that source S\(_i\) is turned on.

\subsection{The effective \(2\times2\) transfer matrix}\label{subsec:effectiveT}

The transfer matrix \(T(z_2,z_1)\) must map the input vector
\(\bm{X}^{(i)}(z_1)\) to the output vector
\(\bm{Y}^{(i)}(z_2)\), i.e.,
\begin{align}
	\label{eq:effectiveTransform}
    T(z_2,z_1)~ \bm{X}^{(i)}(z_1) = \bm{Y}^{(i)}(z_2),
\end{align}
for \(i \in \{1,2\}\), which gives a set of four simultaneous equations
for the four elements of \(T(z_2,z_1)\). To solve the equations, the
double-slit setup is characterized numerically, by calculating
\(\bm{X}^{(i)}(z_1)\) and \(\bm{Y}^{(i)}(z_2)\) for each
source turned on at a time. The transformation equation for both
sources are combined into one matrix equation
\begin{align}
	\label{eq:combinedTransformations}
	T(z_2,z_1) \left(\begin{matrix} \bm{X}^{(1)}(z_1) & \bm{X}^{(2)}(z_1) \end{matrix} \right) =& \left(\begin{matrix} \bm{Y}^{(1)}(z_2) & \bm{Y}^{(2)}(z_2) \end{matrix} \right),
\end{align}
where
\(\left(\begin{matrix} \bm{X}^{(1)}(z_1) & \bm{X}^{(2)}(z_1) \end{matrix} \right)\)
is a $2 \times 2$ matrix with \(\bm{X}^{(1)}(z_1)\) and
\(\bm{X}^{(2)}(z_1)\) as columns, and similar for the right-hand side.
Inverting
Eq.~(\ref{eq:combinedTransformations})
yields
\begin{align}
	\label{eq:TSolved}
	T(z_2,z_1) =  \left(\begin{matrix} \bm{Y}^{(1)}(z_2) & \bm{Y}^{(2)}(z_2) \end{matrix} \right)~  \left(\begin{matrix} \bm{X}^{(1)}(z_1) & \bm{X}^{(2)}(z_1) \end{matrix} \right)^{-1},
\end{align}
provided that
\(\left(\begin{matrix} \bm{X}^{(1)}(z_1) & \bm{X}^{(2)}(z_1) \end{matrix} \right)\)
is invertible. In general,
\(\left(\begin{matrix} \bm{X}^{(1)}(z_1) & \bm{X}^{(2)}(z_1) \end{matrix} \right)\)
is a symmetric matrix because of the symmetry in the setup, and the
diagonal elements are slightly different from the off-diagonal elements
as the projections of field from one source is not equal on both the
post-selected modes. Such a matrix is always invertible.

However, the effective transfer matrix
is not unitary because diffraction is intrinsically a lossy process in which most of the light incident on the slits are blocked by the opaque areas. Moreover, the wavelets, chosen as bases of the Hilbert space of each
  slice, are not eigenfunctions of the Helmholtz equation. Therefore,
  there is cross-talk between different modes as one moves from one
  slice to another. 
  To reveal the underlying unitary transformation, a polar decomposition
of the transfer matrix is done. Such a decomposition factorizes the
non-unitary transfer matrix into a unitary matrix and a Hermitian
matrix.

\subsection{Verifying the double-slit beam splitter}
To check the efficacy of this beam splitter, we study the cross-correlation of the outputs. As the solutions of the Helmholtz equation are time-independent, the cross-correlation is modified such that the distinguishing parameter between the inputs is the angle of polarization \cite{Chiao1992} instead of the time-delay. Further, we concatenate two such double-slit beam splitters to construct a Mach-Zehnder interferometer.

\subsubsection{Cross-correlation of the post-selected output fields}
 Let the field from source $\mathrm{S}_2$ have a phase $\varphi$ with respect to that from source $\mathrm{S}_1$. Also, we rotate the polarization of source S$_2$ so that $\theta$ is the angle between the directions of polarizations of the fields from the two sources. When both the sources are turned on, port $\mathrm{D}_1$ intercepts the vector superposition of fields from sources $\mathrm{S}_1$ and $\mathrm{S}_2$ through slit $\mathrm{A}_1$. The integrated intensity in the post-selected mode on port $\mathrm{D}_1$, i.e., $e_1(x,y;z_2)$ is the projection of the vector sum of the fields intercepted by the port, i.e.,
\begin{align}
	\label{eq:intensityOnD1bothS}
	\left\|E_+(z_2,\theta, \varphi) \right\|^2 \coloneqq &\iint\limits_{z_2} \mathrm{d}x~ \mathrm{d}y~\left|E_+(x,y,z_2,\theta, \varphi) \right|^2\nonumber \\ 
	= &\left|\left\langle e_1(z_2),E^{(1)}_1(z_2) \right\rangle\right|^2 + \left|\left\langle e_1(z_2),E^{(2)}_1(z_2) \right\rangle\right|^2 \nonumber \\
	&+ 2 \mathrm{Re} \left\{\left\langle e_1(z_2),E^{(1)}_1(z_2) \right\rangle^* \left\langle e_1(z_2),E^{(2)}_1(z_2) \right\rangle \right\} \cos\varphi \cos\theta,
\end{align}
where $\theta$ is the distinguishability parameter between the two sources. Similarly, the integrated intensity in the post-selected mode on port $\mathrm{D}_2$ is calculated by projecting $E_-(x,y;z_2,\theta, \varphi)$ on to $e_2(x,y;z_2)$.

The cross-correlation between the two outputs as a function of $\theta$ in Eq.~(\ref{eq:crossCorrelationTheoryExpressionTheta}) is modified for time-independent fields as
\begin{align}
	\label{eq:crossCorrelationTheoryExpressionThetaSlice}
	C(\theta; z_2) \coloneqq \frac{
		\int\limits \mathrm{d} \varphi~ p(\varphi) \left\|E_+(z_2,\theta, \varphi) \right\|^2 \left\|E_-(z_2,\theta, \varphi) \right\|^2
	}
	{
		\int\limits \mathrm{d} \varphi~ p(\varphi) \left\|E_+(z_2,\theta, \varphi) \right\|^2 \int\limits \mathrm{d} \varphi~ p(\varphi) \left\|E_-(z_2,\theta, \varphi) \right\|^2
	}
\end{align}
which is the intensity-intensity correlation of the two output modes of the slice at $z=z_2$.

\subsubsection{Effective Mach-Zehnder interferometer}
\label{sec:MZ}
\begin{figure}[H]
	\centering
	\includegraphics[width=0.7\textwidth]{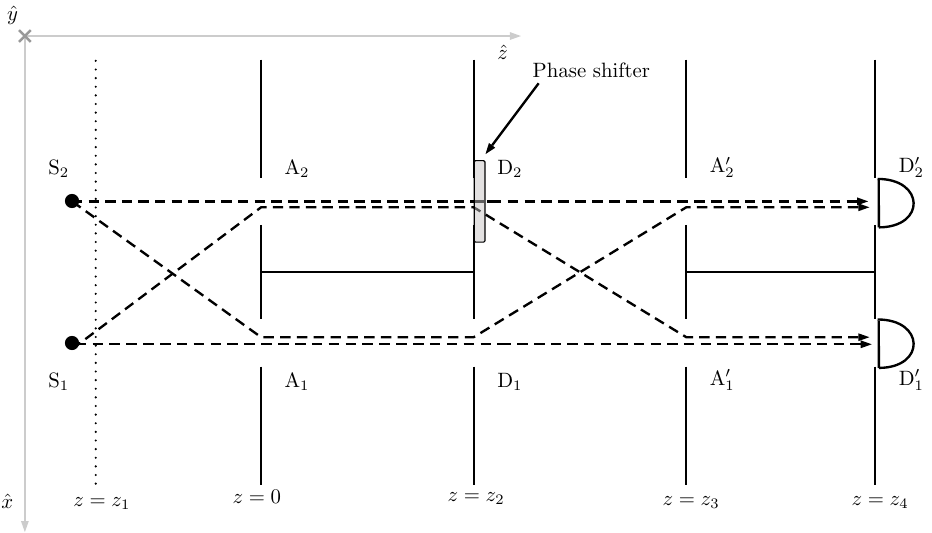}
	\caption{Using two double-slit setups a Mach-Zehnder interferometer is constructed by concatenating them such that both are aligned center-to-center and parallel to each other. The output ports $\mathrm{D}_1$ and $\mathrm{D}_2$ of the first double-slit beam splitter serve as the inputs for the second one. The two detectors are placed behind the output ports D$'_1$ and D$'_2$ of the second double-slit setup. A phase shifter is placed at the output port D$_2$ which changes the phase of the field from that port by $\alpha$.}
	\label{fig:mziSetup}
\end{figure}

The double-slit beam splitters, discussed in this work, can be concatenated to construct more sophisticated interferometers. As an example, Fig.~\ref{fig:mziSetup} shows the schematic of an effective Mach-Zehnder interferometer (MZI) \cite{hariharan2010basics} made by concatenating two such double-slit modules.

The detectors behind ports $\mathrm{D}_1$ and $\mathrm{D}_2$ as in Fig.~\ref{fig:doubleSlitSetup} are removed, and these ports now serve as inputs to the second double-slit module. The fields from these ports get diffracted by slits $\mathrm{A}'_1$ and $\mathrm{A}'_2$ and reach the output ports $\mathrm{D}'_1$ and $\mathrm{D}'_2$, of the second double-slit module behind each of which is a detector. 

A phase shifter (see appendix \ref{append:phaseImplement} on how the phase-shifter is implemented numerically) causes an interference pattern at the output ports $\mathrm{D}'_1$ and $\mathrm{D}'_2$ as the phase, say $\alpha$ is changed in one of the arms of the MZI. The interfernce pattern obtained is used as a signature to verify the double-slit based MZI.

The MZI is essentially a concatenation of two beam splitters. If both the beam splitters are identical 50:50 splitters with transfer matrix in Eq.~(\ref{eq:bsMatrix}) and the phase in one arm, say $\alpha$ is set to zero (the arm lengths are considered equal), the transfer matrix for the MZI is~\cite{hariharan2010basics} 
\begin{align}
\label{eq:mziT}
\frac{1}{2} \trans{1}{\mathrm{i}}{\mathrm{i}}{1} \trans{1}{\mathrm{i}}{\mathrm{i}}{1} = \trans{0}{i}{i}{0},
\end{align}
and therefore the transfer matrix for the double-slit MZI should be close to this. We use a similar approach as that used for the double-slit setup, to find the transfer matrix for the effective MZI, with the output slice at $z_4$ (as shown in Fig.~\ref{fig:mziSetup}).

\subsection{Extending to three dimensions}\label{subsec:tripleSlitModes}
\begin{figure}[H]
    \centering
    \includegraphics[width=0.5\textwidth]{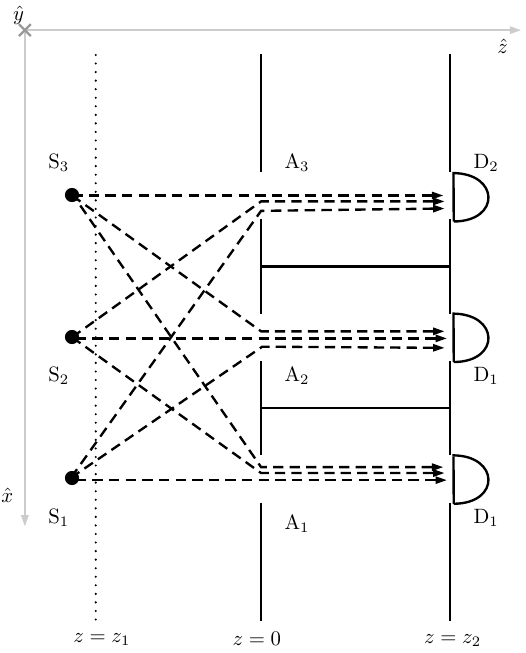}
    \caption{Schematic of a triple-slit setup constructed in a similar way as the double-slit setup in Fig.~\ref{fig:doubleSlitSetup}, by adding a source, slit and detector to the latter.}
    \label{fig:tripleSlitSetup}
\end{figure}

A beam splitter is a two-input-two-output device which, as discussed above, can be effectively constructed using double-slit diffraction. However, one of the key potential uses of slit-diffraction and the framework outlined in this work is extension to higher dimensions. As an example, Fig.~\ref{fig:tripleSlitSetup} shows a triple-slit setup in which a third source S$_3$ is placed at $(-3d/2,0, -L)$, a slit A$_3$ centered at $(-3d/2,0, 0)$ and a detector D$_3$ centered at $(-3d/2,0, L)$. Similar to the double-slit case, the transfer matrix approach can be applied to this system. 

For the triple-slit setup, three Haar scaling functions are chosen as post-selected input modes and three for the post-selected output modes. According to the positions of the detectors (and slits) these modes are
\begin{align}
	\label{eq:3PSInputMode1}
    e_1(x,y;z_1) \coloneqq & g_1(x,y;z_1,-2,2,0),\\
    \label{eq:3PSInputMode2}
    e_2(x,y;z_1) \coloneqq & g_1(x,y;z_1,-2,-3,0),\\
    \label{eq:3PSInputMode3}
    e_3(x,y;z_1) \coloneqq & g_1(x,y;z_1,-2,-8,0),
\end{align}
and similarly for slice at $z_2$. 

Like in the case of two slits, the post-selected input and output will have a 3-dimensional column representation similar to those in Eqs.~(\ref{eq:outputFromSi}) and (\ref{eq:inputFromSi}). The equation for the effective transfer matrix is 
\begin{align}
	\label{eq:3combinedTransformations}
	T(z_2,z_1) \left(\begin{matrix} \bm{X}^{(1)}(z_1) & \bm{X}^{(2)}(z_1) & \bm{X}^{(3)}(z_1) \end{matrix} \right) =& \left(\begin{matrix} \bm{Y}^{(1)}(z_2) & \bm{Y}^{(2)}(z_2) & \bm{Y}^{(3)}(z_2)\end{matrix} \right),
\end{align}
where the superscripts denote the source that is turned on.

\section{Results}
Numerical calculations of the solutions of the Helmholtz equation (with Fraunhofer approximations, see appendix \ref{append:Fraunhofer}) gives the resultant transfer matrices for the double and triple-slit setups. We also show the results of the variation of the correlation of the post-selected outputs of the double-slit setup. Further, we show the numerically calculated interference pattern at the two outputs of the MZI made by concatenating two double-slit beam splitters, and also find its transfer matrix.

\subsection{Effective transfer matrix for the double-slit setup}
As discussed in \S\ref{subsec:effectiveT}, the effective $2 \times 2$ transfer matrix is calculated by characterizing the double-slit setup by turning one source on at a time. For each source the input and output slices have a post-selected vector representation. For the double-slit setup under consideration, the input and output vectors when source S$_1$ is on are
\begin{align}
	\label{eq:inCol1}
	\bm{X}^{(1)}(z_1) \approx &\colvec{-394.761 - 41.473~ \mathrm{i}}{10.284 - 13.398~ \mathrm{i}},\\
    \label{eq:outCol1}
	\bm{Y}^{(1)}(z_2) \approx & \colvec{0.008 - 0.796~\mathrm{i}}{0.782 + 0.008~\mathrm{i}},
\end{align}
with respect to the post-selected modes (see \S\ref{subsec:post-selectedModes}). Similarly, the post-selected vector representations of the input and output slice when source S$_2$ is on are
\begin{align}
	\label{eq:inCol2}
	\bm{X}^{(2)}(z_1) \approx & \colvec{10.284 - 13.398~ \mathrm{i}}{-394.761 - 41.473~ \mathrm{i}},\\
	\label{eq:outCol2}
	\bm{Y}^{(2)}(z_2) \approx & \colvec{0.782 + 0.008~\mathrm{i}}{0.008 - 0.796~\mathrm{i}}.
\end{align}
We use the above results in Eq.~(\ref{eq:TSolved}) to calculate the effective transfer matrix.

With respect to the post-selected input and output vectors, the effective transfer matrix is
\begin{align}
	\label{eq:resultT}
	T(z_2,z_1) \approx &~ \mathrm{e}^{0.476 \pi \mathrm{i}}\trans{2.07}{1.90~ \mathrm{e}^{0.486 \pi \mathrm{i}}}{1.90 ~ \mathrm{e}^{0.486 \pi \mathrm{i}}}{2.07} \times 10^{-3},
\end{align}
which is a symmetric matrix as expected from the symmetry of
the double-slit setup
(Fig.~\ref{fig:doubleSlitSetup}). Note
that apart from a factor of about
\(\mathrm{e}^{0.476 \pi \mathrm{i}} \times 2\sqrt{2} \times 10^{-3}\),
the matrix \(T(z_2,z_1)\) is approximately (but not exactly) a 50:50
beam splitter matrix. The deviation from the ideal 50:50 beam splitter
is due to the non-unitary nature of the transfer matrix. To reveal the
exact unitary transformation the polar decomposition of the transfer
matrix is performed.

\subsection{The underlying unitary transformation}\label{sec:UTransform}
As expected, the transfer matrix \(T(z_2,z_1)\) is not unitary as can be
seen from
\begin{align}
	\label{eq:TTdagger}
	T(z_2,z_1) T^\dagger(z_2,z_1) \approx \trans{7.90}{0.35}{0.35}{7.90} \times 10^{-6} \neq \trans{1}{0}{0}{1},
\end{align}
because of reasons discussed in \S\ref{subsec:effectiveT}. To reveal the underlying unitary
transformation a polar decomposition of \(T(z_2,z_1)\) is carried out
which yields,
\begin{align}
	\label{eq:polarDecomposition}
	T(z_2,z_1) = U(z_2,z_1) P(z_2,z_1),
\end{align}
where the result for the double-slit system considered in this paper, i.e. the polar decompostion of the transfer matrix in Eq.~(\ref{eq:resultT}) is
\begin{align}
\label{eq:unitaryPart}
	U(z_2,z_1) \approx&~ \mathrm{e}^{0.47 \pi \mathrm{i}} \times \frac{1}{\sqrt{2}} \trans{1.04}{0.95~ \mathrm{i}}{0.95~ \mathrm{i}}{1.04},\\
    \label{eq:hermitianPart}
	P(z_2,z_1) \approx& \trans{2.81}{0.06}{0.06}{2.81} \times 10^{-3},
\end{align}
where \(U\) is the transformation of a 54:46 beam splitter upto a global
phase (which is irrelevant as the detectors are square-law type). The
Hermitian component \(P(z_2,z_1)\) captures the non-unitarity of the
transfer matrix. Its diagonal elements show the fraction of the input
that is detected by the detectors after post-selection. The off-diagonal
terms show cross-talk between the two modes.

Therefore, the double-slit setup in
Fig.~\ref{fig:doubleSlitSetup} is
effectively a lossy beam splitter with respect to the post-selected
input and output modes. To verify this result, cross-correlation of
the post-selected outputs is calculated and the result is compared with
what is expected from an ideal beam splitter
(See appendix \ref{append:corr}).

\subsection{Cross-correlation of the post-selected outputs}
\label{subsec:calCorrelation}
\begin{figure}[H]
	\centering
	\begin{minipage}{0.45\textwidth}
		\includegraphics[width=1\textwidth]{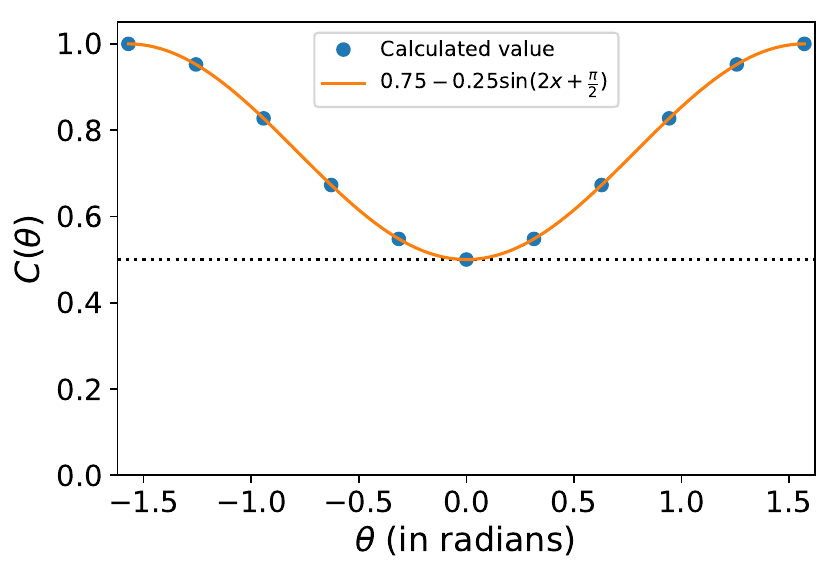}
		\caption{The intensity-intensity correlation of the output in the double-slit setup is calculated using Eq.~(\ref{eq:crossCorrelationTheoryExpressionThetaSlice}) with $\varphi$ chosen from the probability distribution in Eq.~(\ref{eq:pFor50}). As is the case with a regular cubic 50:50 beam splitter, the correlation shows a 50\% dip. The minimum is for $\theta=0$ when both the inputs are indistinguishable, and maximum for $\theta=\pi/2$ when they are completely distinguishable. Compare this with Fig.~\ref{fig:bsCorrelation}.}
		\label{fig:correlation50}
	\end{minipage}\hfill
	\begin{minipage}{0.45\textwidth}
		\includegraphics[width=1\textwidth]{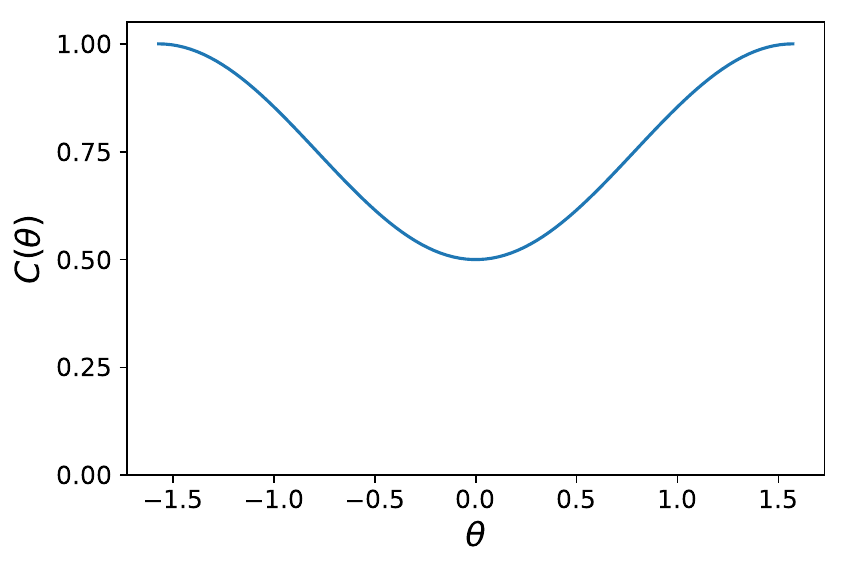}
	\caption{The cross-correlation function plotted as a function of the relative polarization angle $\theta$ between the input pulses, for a cubic beam splitter. The correlation is minimum when both the sources are indistinguishable, i.e., $\theta=0$, and maximum when they are completely distinguishable, i.e., $\theta=\pi/2$. The detailed calculations are presented in appendix \ref{append:corr}}
	\label{fig:bsCorrelation}
	\end{minipage}
\end{figure}

Here we show the result of the numerically calculated intensity-intensity cross-correlation of the post-selected outputs using Eq.~(\ref{eq:crossCorrelationTheoryExpressionThetaSlice}) with 
\begin{align}
    p(\varphi) = \frac{1}{2 \pi},
\end{align}
i.e., the relative phase between the two sources is uniformly random. Figure \ref{fig:correlation50}
shows the values of the correlation as a function of the distinguishability parameter, which in this case is the relative polarization angle $\theta$ between the two sources. The function that fits the result is
\begin{align}
	\label{eq:fit50}
	C_{50}(\theta) = 0.75 - 0.25~ \cos 2\theta,
\end{align} 
the visibility of which is $0.5$.

On the other hand, for a 50:50 cubic beam splitter (with transfer matrix as in Eq.~(\ref{eq:bsMatrix})), if the relative phase $\varphi$ between the two sources is distributed uniformly over the interval $[0, 2\pi)$, the cross-correlation function shows a visibility of $0.5$ as shown in Fig.~\ref{fig:bsCorrelation} (see appendix \ref{append:corr} for more details on the correlation of the outputs of a cubic beam splitter). On comparing the variation of the correlation of the outputs of the double-slit setup with that of the cubic beam splitter, we confirm the beam splitter like behaviour of the double-slit setup.

For completeness, appendix \ref{append:100dip} discusses the 100\% dip in the correlation by using a suitable probability distribution of phase, as suggested in \cite{sadana2018near}.

\subsection{The effective MZI}
\begin{figure}[H]
	\centering
	\includegraphics[width=0.5\textwidth]{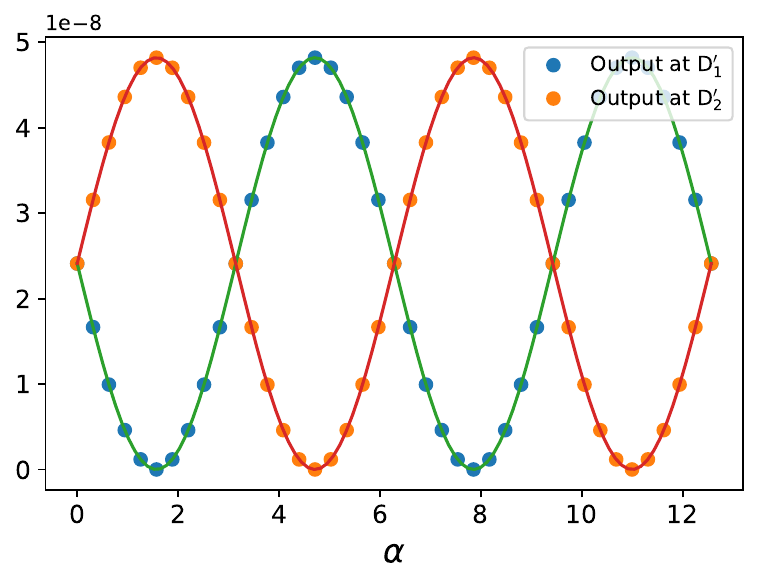}
	\caption{Interference pattern at the output of the MZI made of concatenated double-slit beam splitters.}
	\label{fig:mziInterference}
\end{figure}

Here we show the interference pattern at the output of the double-slit based MZI as shown in Fig.~\ref{fig:mziSetup}. Similar to the case of one double-slit setup, adopting the Fraunhofer approximation and calculating the integrated intensities at the two detectors for different values of $\alpha$ yields an interference pattern shown in Fig.~\ref{fig:mziInterference}. Such an interference pattern is a signature of an MZI. 
The curves that fit the resultant integrated intensities at ports D$'_1$ and D$'_2$ are approximately
\begin{align}
	\label{eq:fitmzi1}
	I_1(\alpha) = 2.41 \times 10^{-8} \left(1 - \sin \left(x - 2.45 \times 10^{-4}\right) \right),\\
	\label{eq:fitmzi2}
	I_2(\alpha) = 2.41 \times 10^{-8} \left(1 + \sin \left(x + 2.45 \times 10^{-4}\right) \right),
\end{align}
respectively. The visibility $V_\mathrm{MZI}$ of both the curves is
\begin{align}
	\label{eq:Vmzi}
	V = 99.94 \%,
\end{align}
which means that the effective MZI using double-slit modules closely emulates an MZI with cubic beam splitters. Therefore, the transfer matrix formalism applied to the setup in Fig.~\ref{fig:mziSetup} should yield the transformation in Eq.~(\ref{eq:mziT}).

Using the method outlined in this paper, the transfer matrix for the double-slit Mach-Zehnder in Fig.~\ref{fig:mziSetup} is
\begin{align}
T_{\mathrm{MZ}} \approx \mathrm{e}^{0.08 \pi \mathrm{i}}\trans{0.255}{3.903~ \mathrm{e}^{0.39 \pi \mathrm{i}}}{3.903~\mathrm{e}^{0.39 \pi \mathrm{i}}}{0.255} \times 10^{-7},
\end{align}
which after polar decomposition yields
\begin{align}
\label{eq:mziU}
U_{\mathrm{MZ}} \approx &\trans{0.061}{0.998~ \mathrm{i}}{0.998~ \mathrm{i}}{0.061} \times \mathrm{e}^{-0.03 \pi \mathrm{i}},\\
\label{eq:mziP}
P_{\mathrm{MZ}} \approx & \trans{3.910}{0.088}{0.088}{3.910} \times 10^{-7}.
\end{align}
Up to a global phase the transfer matrix method successfully reveals the underlying unitary operator for the double-slit Mach-Zehnder which can be checked by comparing Eqs.~(\ref{eq:mziT}) and (\ref{eq:mziU}).

\subsection{Transfer matrix for the triple-slit setup}
We have verified our formalism using a double-slit setup and by comparing it to a well-known optical device, the cubic beam splitter. Here, we apply the transfer matrix formalism to get the post-selected transfer matrix of a triple-slit setup, demonstrating the extensibiltiy of the framework to higher-dimensional systems. 

The effective $3 \times 3$ transfer matrix for the triple-slit setup in Fig.~\ref{fig:tripleSlitSetup}, with respect to the post-selected modes discussed in \S\ref{subsec:tripleSlitModes} is
\begin{align}
    T_{3}(z_2,z_1) \approx \mathrm{e}^{-\mathrm{i} 0.16 \pi}~
    \left(\begin{matrix}
        0.636 & 0.629~\mathrm{i} & 0.594 \\
        0.626~\mathrm{i} & 0.637 & 0.626~\mathrm{i} \\
        0.594 & 0.629~\mathrm{i} & 0.636
    \end{matrix}\right) \times 10^{-3},
\end{align}
where subscript $3$ denotes that the transfer matrix is for a triple-slit setup. The polar decomposition of the $3 \times 3$ transfer matrix results in
\begin{align}
    U_3(z_2,z_1) \approx& \frac{\mathrm{e}^{-\mathrm{i} 0.15 \pi}}{\sqrt{3}}~
   \left(
        \begin{array}{ccc}
         1.493 & 0.844~\mathrm{i} & -0.242 \\
         0.844~\mathrm{i} & 1.255 & 0.844~\mathrm{i} \\
         -0.242 & 0.844~\mathrm{i} & 1.493 \\
        \end{array}
        \right),\\
    P_3(z_2,z_1) \approx& \left(
    \begin{matrix}
        0.772 & 0.146~\mathrm{i} & 0.730 \\
        -0.146~\mathrm{i} & 1.075 & -0.146~\mathrm{i} \\
        0.730 & 0.146~\mathrm{i} & 0.772
    \end{matrix}
    \right) \times 10^{-3},
\end{align}
which reveals the underlying unitary transformation along with the losses captured by the Hermitian component.

By further increasing the number of slits in a similar fashion, even higher-dimensional transfer matrices can be realized using slit-diffraction. Therefore, unlike with four-port devices like beam splitters, which have to be concatenated to implement higher-dimensional transformations, a single $N$-slit setup can be used for an $N$-dimensional transformation. This way of implementing higher-dimensional transfer matrices should prove to be easier than that using orbital angular momentum of light, because of the practical limits on obtaining high OAM states \cite{Garc_a_Escart_n_2011}.

\section{Conclusion}\label{sec:conclusion}
A post-selected unitary representation of slit-diffraction is achieved by projecting the solutions of the Helmholtz equation on two-dimensional plane and finding a transfer matrix that maps one slice to another. The Haar wavelets and scaling functions are used as orthonormal modes that span the slice on each plane. From the infinite set of modes, two input and two output modes are post-selected depending on the area and position of the detectors.
The non-unitary transfer matrix is polar decomposed to reveal the underlying unitary transformation and a Hermitian component that captures the losses. Using this approach,
a double-slit setup, with appropriate modification, 
is effectively a 54:46 beam splitter.

The beam splitter behaviour is verified by calculating the intensity-intensity cross-correlation of the outputs and getting a Hong-Ou-Mandel like variation. Two such double-slit beam splitters are concatenated to construct a Mach-Zehnder interferometer showing that sophisticated interferometers can be constructed using slit based diffraction.
Similar to a double-slit setup,
a higher number of slits can be used to construct a multi-input-multi-output devices, an example of which is shown by finding the post-selected transfer matrix for a triple-slit setup. The future work involves quantizing the fields and making a quantum version of slit-diffraction-based interferometers, which can be used for implementing QIP protocols. 

SS would like to thank A. Sinha for valuable discussions and suggestions; S.N. Sahoo and A. Singh for discussions and proof-reading of the manuscript. SS also thanks the Canadian Queen Elizabeth II Diamond Jubilee Scholarships program (QES) for funding his visit to the University of Calgary during the course of this project. BCS appreciates the VAJRA fellowship support from SERB, Govt.~of India.

\appendixpage
\appendix

\section{Fraunhofer approximation to solutions of Helmholtz equation}\label{append:Fraunhofer}
Here we derive the solutions of the Helmholtz equation when only source S$_1$ is switched on in the double-slit setup in Fig.~\ref{fig:doubleSlitSetup}, i.e.,
\begin{align}
\label{eq:HelmholtzEq}
\left( \nabla^2 + k^2 \right) E(\bm{r}) = \delta^3(\bm{r} - \bm{r}_{\mathrm{S}_1}),
\end{align}
with the boundary conditions specified in appendix \ref{append:slice2slice}. Here we have ignored the $10^5$ factor that is the amplitude of the source, as it will only re-scale the diffracted field by that factor. The first subsection considers the far-field approximation and the second subsection considers the slit width to be very small compared to the distance between the center of the slit and point at which the field is calculated. 

\subsection{Far-field approximations applied to the propagator}
The complete solution of the Helmholtz equation (Eq.~(\ref{eq:HelmholtzEq})) in a volume $\mathcal{V}$ enclosed by a surface $\partial \mathcal{V}$ is 
\begin{align}
    \label{eq:genSolSpecific}
    E(\bm{r}) = &\int\limits_\mathcal{V} \text{d}^3 \bm{r}' G(\bm{r},\bm{r}')~ \delta^3(\bm{r}'-\bm{r}_{\mathrm{S}_1})
    + \int\limits_\mathcal{\partial \mathcal{V}} \text{d}^2 \bm{r}'~ \hat{\bm{n}}(\bm{r}')\cdot\left(E(\bm{r}') \nabla'G(\bm{r},\bm{r}')- G(\bm{r},\bm{r}') \nabla'E(\bm{r}')\right).
\end{align}
Consider the surface $\partial \mathcal{V}$ to be at infinity such that there is no contribution from the surface term in Eq.~(\ref{eq:genSolSpecific}). As we have assumed that the opaque portions of the slit-plane are perfectly absorbing, the field within the area of the slit is
\begin{equation}
    E(\bm{r}) = G(\bm{r},\bm{r}_f).
\end{equation}
and zero outside the area of the slit. Here 
\begin{align}
    \label{eq:greensFunction}
    G(\bm{r},\bm{r}') = -\frac{1}{4 \pi}~ \frac{\E^{\I k \|\bm{r}-\bm{r}'\|}}{\|\bm{r}-\bm{r}'\|}
\end{align}
is the Green's function of the Helmholtz equation in three dimension. To find the diffracted field from the slit, consider a semi-infinite volume with the slit-plane as one part of the surface $\partial \mathcal{V}$ and the other parts at infinity as discussed in appendix \ref{append:slice2slice}. This volume does not contain any source, but the surface on the slit-plane gives contribution from the slits. Therefore, we use the surface term of Eq.~(\ref{eq:genSolSpecific}) to calculate the diffracted field as 
\begin{equation}\label{eq:surfaceField}
     \int\limits_{z = \bm{R}_s\cdot \hat{\bm{z}}} \text{d}^2 \bm{r}'~ \hat{\bm{z}}\cdot\left(G(\bm{r}',\bm{r}_f) \nabla'G(\bm{r},\bm{r}') - G(\bm{r},\bm{r}') \nabla'G(\bm{r}',\bm{r}_f)\right),
\end{equation}
where the only contribution is from the slits-plane. To further simplify the expression for the diffracted field, note that
\begin{align}
    \nabla_2 G(\bm{r}_2,\bm{r}_1) &= -\frac{1}{4\pi}~ \frac{\I k \E^{\I k |\bm{r}_2-\bm{r}_1|}}{|\bm{r}_2-\bm{r}_1|} - \frac{\E^{\I k |\bm{r}_2-\bm{r}_1|}}{|\bm{r}_2-\bm{r}_1|^2}~ \nabla_2 |\bm{r}_2-\bm{r}_1| \nonumber \\
    &= -\frac{1}{4\pi}~ \frac{\E^{\I k |\bm{r}_2-\bm{r}_1|}}{|\bm{r}_2-\bm{r}_1|} 
    \left(\I k  - \frac{1}{|\bm{r}_2-\bm{r}_1|}\right)~ 
    \frac{\bm{r}_2-\bm{r}_1}{|\bm{r}_2-\bm{r}_1|},
\end{align}
which means that
$\nabla_1 G(\bm{r}_2, \bm{r}_1) = - \nabla_2 G(\bm{r}_2, \bm{r}_1)$ and from Eq.~(\ref{eq:greensFunction})  $G(\bm{r}_2,\bm{r}_1) = G(\bm{r}_1,\bm{r}_2)$. The far-field is applied by assuming that $|\I k| >> \frac{1}{|\bm{r}_2-\bm{r}_1|}$. Then 
\begin{equation}
    \nabla_2 G(\bm{r}_2,\bm{r}_1) \approx \I k~ G(\bm{r}_2,\bm{r}_1)~ \frac{\bm{r}_2-\bm{r}_1}{|\bm{r}_2-\bm{r}_1|}.
\end{equation}
With this approximation, Eq.~(\ref{eq:surfaceField}) simplifies to
\begin{align}\label{eq:farfield}
    E(\bm{r}) &= -\I k \int\limits_\mathcal{S} \text{d}^2 \bm{r}'~ 
    G(\bm{r}',\bm{r}_f) G(\bm{r},\bm{r}')~
    \left(\frac{z-z'}{|\bm{r}-\bm{r}'|} + \frac{z'-z_f}{|\bm{r}'-\bm{r}_f|}\right).
\end{align}

\subsection{Small slit approximation}
The expression for the diffracted field simplifies further when the small-slit approximation is considered. Let $\bm{R}_s$ be the center of the slit with a width very small compared with its distance from the point of detection. A position within the area of the slit can be written as $\bm{r}' = \bm{R}_s + \bm{\Delta}_s$. Therefore,
\begin{align}
    G(\bm{r}',\bm{r}_f) &= -\frac{1}{4\pi}~ \frac{\E^{\I k |\bm{R}_s-\bm{r}_f + \bm{\Delta}_s|}}{|\bm{R}_s-\bm{r}_f + \bm{\Delta}_s|},
\end{align}
where the small slit approximation is applied as $|\bm{\Delta_s}| / |\bm{R}_s - \bm{r}_f| \ll 1$ so that
\begin{align}
    |\bm{R}_s - \bm{r}_f + \bm{\Delta}_s| &= \sqrt{|\bm{R}_s-\bm{r}_f|^2 + |\bm{\Delta}_s|^2 + 2 \bm{\Delta}_s\cdot(\bm{R}_s-\bm{r}_f)} \nonumber \\
    &= |\bm{R}_s-\bm{r}_f| \sqrt{1 + \frac{|\bm{\Delta}_s|^2}{|\bm{R}_s-\bm{r}_f|^2} + 2 \bm{\Delta}_s\cdot\frac{(\bm{R}_s-\bm{r}_f)}{|\bm{R}_s-\bm{r}_f|^2}} \nonumber \\
    &\approx |\bm{R}_s-\bm{r}_f| \sqrt{1 + 2 \bm{\Delta}_s\cdot\frac{(\bm{R}_s-\bm{r}_f)}{|\bm{R}_s-\bm{r}_f|^2}} \nonumber \\
    &\approx |\bm{R}_s-\bm{r}_f| \left(1 + \bm{\Delta}_s\cdot\frac{(\bm{R}_s-\bm{r}_f)}{|\bm{R}_s-\bm{r}_f|^2}\right).
\end{align}
Furthermore, for far-field and small slit
\begin{align}
    \frac{1}{|\bm{r}'-\bm{r}_f|} \approx \frac{1}{|\bm{R}_s-\bm{r}_f|}
\end{align}
which results in 
\begin{align}
    G(\bm{r}',\bm{r}_f) &= -\frac{1}{4\pi}~ \frac{\E^{\I k |\bm{R}_s-\bm{r}_f|}}{|\bm{R}_s-\bm{r}_f|}~
    \E^{\I k \bm{\Delta}_s \cdot \frac{\bm{R}_s-\bm{r}_f}{|\bm{R}_s-\bm{r}_f|}} \nonumber\\
    &= G(\bm{R}_s,\bm{r}_f)~ \E^{\I k \bm{\Delta}_s \cdot \frac{\bm{R}_s-\bm{r}_f}{|\bm{R}_s-\bm{r}_f|}}.
\end{align}
Similarly,
\begin{align}
    G(\bm{r},\bm{r}') &= G(\bm{r},\bm{R}_s)~ \E^{-\I k \bm{\Delta}_s \cdot \frac{\bm{r}-\bm{R}_s}{|\bm{r}-\bm{R}_s|}},
\end{align}
and using this approximation in Eq.~(\ref{eq:farfield}) we get 
\begin{align}\label{eq:farSmallSlit}
    E(\bm{r}) =&-\I k \int\limits_\mathcal{S} \text{d}^2 \bm{\Delta}_s~ 
   G(\bm{r},\bm{R}_s)~ G(\bm{R}_s,\bm{r}_f)~
   \E^{-\I k \bm{\Delta}_s \cdot \left(\frac{\bm{r}-\bm{R}_s}{|\bm{r}-\bm{R}_s|}-\frac{\bm{R}_s-\bm{r}_f}{|\bm{R}_s-\bm{r}_f|}\right)}~ \nonumber\\
    &\times \left(\frac{z-z_s}{|\bm{r}-\bm{R}_s|} + \frac{z_s-z_f}{|\bm{R}_s-\bm{r}_f|}\right).
\end{align}
For a rectangular slit $w_x$ wide along $x$ and $w_y$ along $y$, the integral is a $\mathrm{sinc}$ function and the final expression for the field of the slice is
\begin{align}
    E(\bm{r}) = &-\I k w_x w_y~
    \left(\frac{z-z_s}{|\bm{r}-\bm{R}_s|} + \frac{z_s-z_f}{|\bm{R}_s-\bm{r}_f|}\right)~
    G(\bm{r},\bm{R}_s)~ G(\bm{R}_s,\bm{r}_f)~\nonumber\\
    &\times \mathrm{sinc}\left(\frac{k w_x}{2}\left(\frac{\bm{r}-\bm{R}_s}{|\bm{r}-\bm{R}_s|}-\frac{\bm{R}_s-\bm{r}_f}{|\bm{R}_s-\bm{r}_f|}\right)\cdot\hat{\bm{x}}\right)~\nonumber\\
    &\times \mathrm{sinc}\left(\frac{k w_y}{2}\left(\frac{\bm{r}-\bm{R}_s}{|\bm{r}-\bm{R}_s|}-\frac{\bm{R}_s-\bm{r}_f}{|\bm{R}_s-\bm{r}_f|}\right)\cdot\hat{\bm{y}}\right),
\end{align}
which with appropriate position vectors of the sources and slit in the double-slit setup yields the diffracted field in that system.

\section{Slice-to-slice map}\label{append:slice2slice}
The surface term of the formal solution of the Helmholtz equation is used to find the slice-to-slice map as follows. Consider the double-slit setup shown in Fig.~\ref{fig:doubleSlitSetup} with only source S$_1$ switched on. The Helmholtz equation for the field within this boundary is Eq.~(\ref{eq:HelmholtzEq}). As the slits are far from the source, the surface term of the formal solution of the Helmholtz equation in Eq.~(\ref{eq:genSolHelmholtz}) can be dropped and the field on the plane $z=z_1$ is approximately $G(\bm{r}, \bm{r}_{\mathrm{S}_1})$ (discussed in the main text in Eq.~(\ref{eq:sourceField})) where $\bm{r}\cdot z = z_1$, i.e., the field is projected on this plane and hence is a slice of the 3D solution.

The field at another plane, say $z=z_2$ can be calculated from the slice at $z_1$. For this, consider a semi-infinite volume enclosed by the planes $z=z_1$, $z \rightarrow \infty$, $x \rightarrow -\infty$, $x \rightarrow \infty$, $y \rightarrow -\infty$ and $y \rightarrow \infty$. This volume includes the double-slit, the source is now excluded, leading to a homogeneous Helmholtz equation within the volume, albeit with complications due to the presence of the slits, whose opaque parts will have some dielectric constant other than one. Because of this the Green's function, say $\tilde{G}(\bm{r},\bm{r}')$ is no longer the free-space Green's function, near the slit plane.

However, this approach yields the slice-to-slice map directly, as the solution at any point within the volume will have contribution only from the slice at $z_1$ because the other surfaces are at infinity. Therefore if one defines a propagator 
\begin{align}
    \label{eq:slice2sliceProp}
    \mathcal{P}(\bm{r},\bm{r}') \coloneqq &  \hat{\bm{z}} \cdot \left(E(\bm{r}') \nabla' \tilde{G}(\bm{r},\bm{r}') - \tilde{G}(\bm{r},\bm{r}') \nabla' E(\bm{r}') \right),
\end{align}
the field within the volume can be calculated from the slice at $z_1$ as
\begin{align}
    E(\bm{r}) =  \iint\limits_{z=z_1} \mathrm{d}^2 \bm{r}'~ \mathcal{P}(\bm{r},\bm{r}')~ E(\bm{r}'),
\end{align}
where the integration is over the plane $z=z_1$. Finally, the slice at $z_2$ is the projection of the field on the plane $z=z_2$, i.e., $\left. E(\bm{r}) \right|_{z_2}$.

The basis constructed using Haar scaling functions and the wavelet functions form a discrete orthonormal basis for the slices. This facilitates a matrix representation of the propagator that maps one slice to another. For example, the matrix representation of the free slice-to-slice propagator $\mathcal{P}(\bm{r}_\perp,\bm{r}'_\perp; z,z')$ in the new basis is
\begin{align}
	\centering
    \label{eq:matRepFreeProp}
    \mathcal{P}_{\imath \jmath}(z_2,z_1) = \iint\limits \mathrm{d}^2 \bm{r}_\perp \iint\limits \mathrm{d}^2 \bm{r}'_\perp~ g_\imath(\bm{r}_\perp;z_2)~ \mathcal{P}(\bm{r}_\perp, \bm{r}'_\perp; z_2,z_1)~ g_\jmath(\bm{r}'_\perp;z_1),
\end{align}
where $g_i(\bm{r};z)$ is a basis function on slice at $z$ and $g_j(\bm{r}';z')$ is that on slice at $z'$. Note that the basis is infinite-dimensional and therefore so is the matrix representation of the free propagator.

\section{The cross-correlation of outputs in a beam splitter}\label{append:corr}
In semi-classical theory of photo-detection \cite{Mandel_1958,Mandel_1964,mandel1995optical}, the probability of coincident photo-detections is proportional to the intensity-intensity cross-correlation of the outputs, a normalized version of which is
\begin{align}
	\label{eq:crossCorrelationTheoryExpression}
        C(\tau) := \frac{\int\text{d}\varphi~ p\left(\varphi\right)
	\int\limits_{T_\text{on}}^{T_\text{off}}\text{d}t\left|E_+(t;\tau,\varphi)\right|^2
	\int\limits_{T_\text{on}}^{T_\text{off}}\text{d}t'\left|E_-(t';\tau,\varphi)\right|^2}
	{\left[\int\text{d}\varphi~ p\left(\varphi\right)
	\int\limits_{T_\text{on}}^{T_\text{off}}\text{d}t
	\left|E_+(t;\tau,\varphi)\right|^2
	\right]
	\left[\int\text{d}\varphi'~ p\left(\varphi'\right)
	\int\limits_{T_\text{on}}^{T_\text{off}}\text{d}t
	\left|E_-(t;\tau,\varphi')\right|^2
	\right]},
\end{align}
where $E_+$ and $E_-$ are the output pulses when the input pulses have a time delay $\tau$ between them and a relative phase $\varphi$. The phase $\varphi$ fluctuates with a probability distribution $p(\varphi)$ and, $T_\text{on}$ and $T_\text{off}$ are detector on and off times respectively. The delay $\tau$ plays the role of a distinguishability parameter between the two input pulses. The cross-correlation is a measure of the fourth-order interference between the two outputs. For a 50:50 beam splitter, $C(\tau)$ shows a variation dependent on the shape of the pulse. If the probability distribution $p(\varphi)$ is uniform over the interval $\left[0, 2\pi \right)$, i.e.,
\begin{align}
	\label{eq:pFor50}
	p(\varphi) = \frac{1}{2 \pi},
\end{align}
the curve shows a visibility of $0.5$. A detailed analysis of this cross-correlation with classical pulses is discussed in \cite{sadana2018near}.

If the distinguishability parameter is the angle of polarization $\theta$ between the two input pulses instead of time-delay $\tau$, the cross-correlation of the intensities can be redefined as
\begin{align}
	\label{eq:crossCorrelationTheoryExpressionTheta}
	C(\theta):=\frac{\int\text{d}\varphi~ p\left(\varphi\right)
	\int\limits\text{d}t \left|E_+(t;\theta,\varphi)\right|^2
	\int\limits\text{d}t' \left|E_-(t';\theta,\varphi)\right|^2}
	{\left[\int\text{d}\varphi~ p\left(\varphi\right)
	\int\limits\text{d}t
	\left|E_+(t;\theta,\varphi)\right|^2
	\right]
	\left[\int\text{d}\varphi'~ p\left(\varphi'\right)
	\int\limits\text{d}t
	\left|E_-(t;\theta,\varphi')\right|^2
	\right]},
\end{align}
where the time-delay between the input pulses is zero. For a 50:50 beam splitter, $C(\theta)$ shows a sinusoidal variation as shown in Fig.~\ref{fig:bsCorrelation}, for uniformly randomized phase as in Eq.~(\ref{eq:pFor50}). 
\begin{figure}[H]
	\centering
	\includegraphics[width=0.5\textwidth]{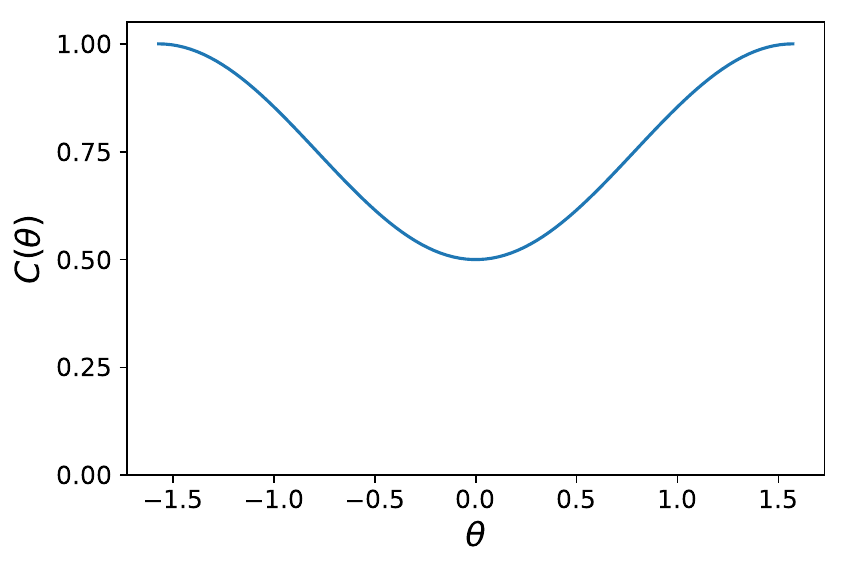}
	\caption{The cross-correlation function plotted as a function of the relative polarization angle $\theta$, which is the distinguishability parameter. The correlation is minimum when both the sources are indistinguishable, i.e., $\theta=0$, and maximum when they are completely distinguishable, i.e., $\theta=\pi/2$. The plot has been generated using Eq.~(\ref{eq:crossCorrelationTheoryExpressionTheta}) for a regular cubic 50:50 beam splitter with two identical input pulses having zero delay between them. When the distinguishability parameter is $\theta$ the shape of the pulses does not affect the correlation.}
	\label{fig:bsCorrelation2}
\end{figure}
The equation that fits the data in the plot of Fig.~\ref{fig:bsCorrelation} is 
\begin{align}
	\label{eq:bsCorrFit}
	C_{\mathrm{fit}}(\theta) = 0.75 - 0.25 \cos 2\theta,
\end{align}
which has a visibility of $0.5$.

As $C(\tau)$ and $C(\theta)$ are dependent on the probability distribution of the phase $\varphi$, the visibility of the curves can exceed $0.5$, with an appropriate choice of the probability distribution. For some distribution, the visibility can reach $1$, classically \cite{sadana2018near}.  The variation of the correlation as a function of the distinguishability parameter is used as a signature of a beam splitter, which the double-slit setup, as is discussed in this work, also exhibits.

\section{The implementation of the phase shifter in MZI}\label{append:phaseImplement}
The phase shifter in the double-slit MZI is modelled as a medium of thickness $t$ and with refractive index $n$. Within the medium the Green's function (and hence the propagator) will change to
\begin{align}
\label{eq:Green'sFunctionWithn}
G(\bm{r},\bm{r}';n) = - \frac{1}{4\pi} \frac{\mathrm{e}^{\mathrm{i} n k \left| \bm{r} - \bm{r}' \right|}}{\left| \bm{r} - \bm{r}' \right|},
\end{align}
where the refractive index of the medium causes a change in the propagation constant resulting in bending of light and a change in the phase. The thickness of the medium is small enough that the effect can be approximated by an extra phase 
\begin{align}
\label{eq:phaseWithn}
\alpha = \frac{2 \pi}{\lambda} (n - 1) t,
\end{align}
imparted to the field and the net effect is captured by simply multiplying the output at port D$_2$ by $\mathrm{e}^{\mathrm{i} \alpha}$.


\section{100\% dip in correlation of the outputs of the double-slit beam splitter}\label{append:100dip}
 
\begin{figure}[H]
		\centering
		\includegraphics[width=0.5\textwidth]{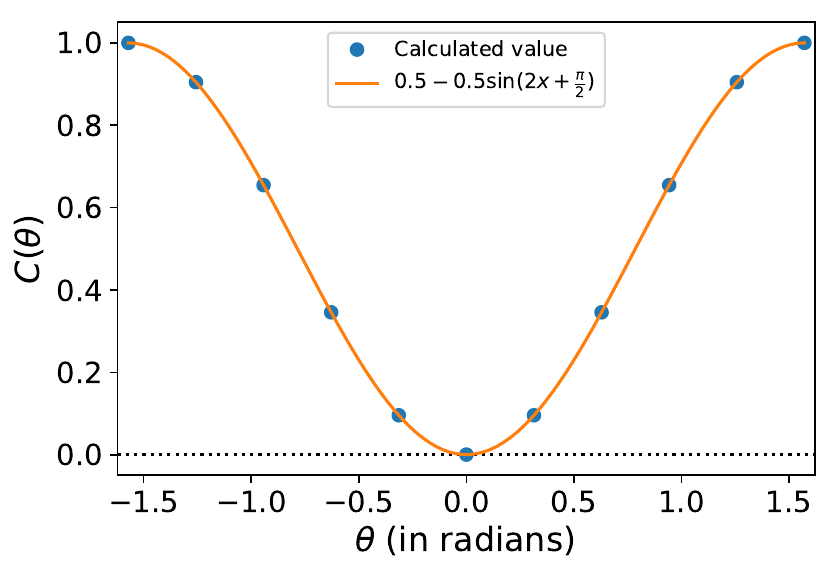}
		\caption{The intensity-intensity correlation of the output in the double-slit setup is calculated using Eq.~(\ref{eq:crossCorrelationTheoryExpressionThetaSlice}) with $\varphi$ chosen from the probability distribution is Eq.~(\ref{eq:pFor100}). In this case the correlation shows a dip of 100\%. Although the fields are classical, a 100\% dip or a Hong-Ou-Mandel like dip is achieved if the probability distribution of the relative phase between the inputs are chosen carefully.}
		\label{fig:correlation100}
	\end{figure}
The visibility of the correlation is dependent on the probability distribution $p(\varphi)$. In particular, if
\begin{align}
	\label{eq:pFor100}
	p(\varphi) = \frac{1}{2} \delta \left(\theta- \frac{\pi}{2} \right) + \frac{1}{2}\delta \left(\theta + \frac{\pi}{2} \right),
\end{align}
the correlation function shows a visibility of $1.0$, as shown in Fig.~\ref{fig:correlation100}.
\noindent The function that fits the result is
\begin{align}
	\label{eq:fit100}
	C_{100}(\theta) = 0.5 - 0.5~ \cos 2\theta,
\end{align} 
which shows that although the fields are classical, a 100\% dip or a Hong-Ou-Mandel like dip is achieved if the probability distribution of the relative phase between the inputs are chosen carefully. A study of such an effect is discussed in \cite{sadana2018near}.

\bibliography{main}
\bibliographystyle{ieeetr}

\end{document}